# ANALISIS KEPUASAN PENGGUNA APLIKASI BINTANG *CASH & CREDIT* MENGGUNAKAN METODE *END USER COMPUTING SATISFACTION* (EUCS)

**Rahayu Agustina[1], Leon Andretti Abdillah[2]**
Fakultas Ilmu Komputer, Universitas Bina Dama
Email: ragustina162@gmail.com[1], leon.abdilah@yahoo.com[2]

*ABSTRACT*

*The use of android application technology has advanced rapidly in recent years, making it one of the alternative media for distributing information in a variety of industries, including e-commerce, that consumers may access at any time and from any location. The Bintang Cash & Credit store in Palembang is one among the stores that has already used the Android application. In EUCS there are seven variables: content, accuracy, format, ease of use and timeliness, security and speed of response. The data of this research were collected by distributing questionnaires to 95 respondents using random sampling technique. Furthermore, the data obtained were processed using SPSS version 25 software. The data analysis method used was quantitative analysis method using validity and reliability tests, classical assumption tests, multiple regression tests and hypothesis testing. From the results of this study, there is a positive influence on the satisfaction of users of the Bintang Cash & Credit application.*

**ABSTRAK**

Penggunaan teknologi aplikasi android saat ini telah berkembang dengan sangat cepat sehingga menjadikan salah satu alternatif media penyampaian informasi disegala bidang termasuk dibidang *e-commerce* yang dapat digunakan tanpa ada batasan waktu maupun tempat oleh penggunanya. Salah satu toko yang sudah menggunakan aplikasi android adalah toko Bintang *Cash & Credit* yang ada di Palembang. Tujuan dari penelitian ini adalah untuk mengetahui sejauh mana tingkat kepuasan pengguna dalam menggunakan aplikasi Bintang *Cash & Credit* menggunakan metode *End User Computing Satisfaction* (*EUCS*). Dalam EUCS terdapat tujuh variabel yaitu, isi (*content*), keakuratan (*accuracy)*, bentuk (*format*), kemudahan penggunaan (*ease of use*) dan ketepatan waktu (*timeliness*), Keamanan (*security*) dan kecepatan (*speed of response*). Data penelitian ini dikumpulkan dengan menyebarkan kuesioner kepada 95 responden dengan menggunakan teknik *sampel randon sampling*. Selanjutnya data yang diperoleh diolah menggunakan software SPSS versi 25. Metode analisa data yang digunakan adalah metode analisa kuantitatif dengan menggunakan uji validitas dan uji reabilitas, uji asumsi klasik, uji regresi berganda dan uji hipotesis. Dari hasil penelitian ini bahwa terdapat pengaruh yang positif terhadap kepuasan penggunaka aplikasi Bintang *Cash & Credit*.

**Kata kunci**: Aplikasi Bintang *Cash & Credit*, kepuasan Pengguna, EUCS

## 1. PENDAHULUAN

Saat ini perkembangan teknologi informasi sangat cepat dan amat pesat sehingga orang-orang dengan cepat mendapatkan informasi hanya dengan hitungan detik saja. Untuk mendapatkan akses informasi yang cepat dan *update* merupakan salah satu tuntutan bagi suatu perusahaan yang mengandalkan penyampaian informasinya melalui internet kepada *customer* maupun masyarakat saat ini. Informasi-informasi tersebut sangatlah mudah didapatkan melalui teknologi jaringan internet yang telah tersebar luas didunia seperti melalui situs web atau aplikasi *android*.

Penggunaan teknologi aplikasi *android* saat ini telah berkembang dengan sangat cepat sehingga menjadikan salah satu alternatif media penyampaian informasi disegala bidang termasuk dibidang *e-commerce* yang dapat digunakan tanpa ada batasan waktu maupun tempat oleh penggunanya. Aplikasi *android* adalah salah satu media yang dapat mewakili sebuah perusahaan atau instansi tertentu untuk menawarkan produk dan jasanya secara cepat dan tepat melalui media aplikasi android [3]. Aplikasi yang baik adalah aplikasi yang dapat memenuhi kepuasan penggunanya terhadap aplikasi tersebut maka untuk mengetahui apakah aplikasi *android* sudah memenuhi kepuasan penggunanya, maka diperlukannya suatu proses analisis dan terdapat beberapa metode yang dapat digunakan, salah satunya adalah metode *End User Computing Satisfaction* (EUCS).

*End User Computing Satisfaction* (EUCS) adalah metode untuk mengukur tingkat kepuasan dari pengguna suatu sistem aplikasi dengan membandingkan antara harapan dan kenyataan dari sebuah sistem informasi [1]. Instrumen EUCS terdiri dari 12 item dengan membandingkan lingkungan pemrosesan data tradisional dengan lingkungan *End User Computing Satisfaction* yang meliputi 5 variabel yaitu isi (*content*), akurasi (*accuracy*), bentuk (*format*), kemudahan pengguna (*ease of use*) dan ketepatan waktu (*timeliness*). Dalam penelitian ini terdapat 2 variabel tambahan yaitu keamanan (*security*) dan kecepatan (*speed of response*) untuk mengukur kepuasan pengguna. Sehingga didapatkan 7 variabel *independen* yang meliputi isi (*content*), akurasi (*accuracy*), bentuk (*format*), kemudahan pengguna (*ease of use*) dan ketepatan waktu (*timeliness*), keamanan (*security*), kecepatan (*speed of response*) dan terdapat 1 variabel *dependen* yaitu kepuasan pengguna akhir (*end user satisfaction*) [8]. Berikut ini adalah indikator setiap variabel:

1) Isi (*content*) terdiri dari 3 indikator yaitu sistem menyediakan informasi yang tepat dan sesuaisistem menyediakan informasi yang lengkap dan informasi yang dihasilkan sistem sangat bermanfaat,
2) Akurasi (*accuracy*) terdiri dari 3 indikator yaitu informasi yang dihasilkan sistem sangat akurat, *output* yang dihasilkan sistem sesuai dengan yang diperintahkan dan sistem menghasilkan informasi yang dapat diandalkan dan dipercaya.
3) Bentuk (*format*) terdiri dari 4 indikator yaitu informasi yang ditampilkan sangat jelas, tampilan *interface* sistem sangat menarik, komposisi warna sistem tidak melelahkan mata dan bentuk laporan yang dihasilkan mudah dimengerti dan dipahami.
4) Kemudahan penggunaan (*ease of use*) teridri dari 3 indikator yaitu sistemnya *user-friendly*, sistem mudah dipahami dan mudah dalam berinteraksi dengan sistem.
5) Ketepatan waktu (*timeliness*) terdiri dari 3 indikator yaitu sistem memberikan informasi tepat waktu, sistem memberikan data terbaru dan sistem menyediakan informasi secara cepat.
6) Keamanan (*security*) terdiri dari 4 indikator yaitu sistem menjamin keamanan informasi agar tidak diakses orang lain, sistem dapat menjamin kerahasiaan informasi, informasi dalam sistem tidak dapat diubah kecuali oleh pemilik informasi dan sistem menyediakan fitur *login* dan *logout*.
7) Kecepatan (*speed of response*) terdiri dari 2 indikator yaitu sistem memililiki kecepatan akses ke *hompage* dan sistem memiliki kecepatan dalm mengakses antar halaman-halaman *web*.
8) Kepuasan pengguna (*user satisfaction*) terdiri dari 3 indikator yaitu sistem yang disediakan dapat membantu tugas atau pekerjaan, sistem efektif dalam penggunaan, sistem efisien dalam penggunaan dan kinerja sistem memenuhi kepuasan penggunanya.

Toko Bintang *Cash & Credit* saat ini sudah memliki aplikasi *android* untuk pembelinya yang diberinama aplikasi Bintang *Cash & Credit*. Aplikasi Bintang *Cash & Credit* adalah suatu aplikasi berbasis *android* yang dibuat untuk memudahkan pembelinya dalam membeli dan mengajukan barang kredit tanpa harus dibatasi ruang dan waktu. Pengguna dari aplikasi Bintang *Cash & Credit* sampai dengan bulan maret 2021 telah diunduh sebanyak 1826 kali. Hal tersebut dapat menjadikan aplikasi Bintang *Cash & Credit* sebagai aplikasi yang berguna bagi penggunanya.

Hal inilah yang membuat peneliti tertarik untuk melakukan penelitian yang berhubungan dengan tingkat kepuasan pengguna aplikasi terutama bagi pengguna yang terlibat dalam penggunaan aplikasi Bintang *Cash & Credit*. Kepuasan pengguna merupakan salah satu indikator dari keberasilan pengembangan sistem informasi. Sistem informasi dapat diandalkan apabila memiliki kualitas yang baik dan mampu memberikan kepuasan pada pemakainya [9].Terdapat beberapa metode yang dapat digunakan, salah satunya adalah metode *End User Computing Satisfaction* (EUCS).

## 2. METODOLOGI PENELITIAN

Objek penelitian ini adalah aplikasi Bintang *Cash & Credit*. Dalam penelitian ini melibatkan 2 (dua) macam metode pengumpulan data yang digunakan [6] yatiu data primer dan data sekunder. Data primer yang diperoleh berupa data hasil kuisioner, observasi dan wawancara, sedangkan data sekunder berupa literatur atau jurnal-jurnal sebagai panduan dalam melaksanakan penelitian.

### 2.1 Populasi dan Teknik Pengambilan Sampel

Populasi adalah wilayah generalisasi yang terdiri atas objek atau subjek yang mempunyai kualitas dan karakteristik tertentu yang ditetapkan oleh peneliti untuk dipelajari dan kemudian ditarik kesimpulannya [7].

Populasi dalam penelitian ini adalah pengguna yang telah mengunduh aplikasi Bintang *Cash & Credit* sebanyak 1826 kali sampai dengan bulan Maret 2021 yang bersumber dari *google playstore*.

Sampel adalah bagian dari jumlah dan karakteristik yang dimiliki oleh populasi tersebut [7]. Penelitian ini menggunakan teknik pengambilan sampel (Sampling) karena peneliti tidak mampu menjangkau keseluruhan populasi. Pada penelitian ini digunakan rumus slovin untuk menentukan sampel minimal dengan menggunakan taraf sgnifikan 10%.

### 2.2 Analisis Kuantitatif

Metode analisis kuantitatif dapat diartikan sebagai metode penelitian yang berlandaskan pada filsafat positivisme, digunakan untuk meneliti pada populasi atau sampel tertentu, pengumpulan data menggunakan instrumen penelitian, analisis data bersifat kuantitatif atau statistik dengan tujuan untuk menguji hipotesis yang telah ditetapkan [7].

Dalam analisis ini menggunakan analisis pearson yang merupakan uji statistik variabel yang berskala interval dimana alternatif jawaban responden berdasarkan kuisioner yang diisi oleh responden yang diberikan bobot berskala likert yaitu 5, 4, 3, 2, 1 untuk setiap pernyataan.

### 2.3 Pengujian Kualitas Data

Pengumpulan data yang dilakukan dalam penelitian ini adalah dengan menggunakan kuesioner, oleh karena itu untuk mengelola datanya maka dilakukan uji validitas dan uji reliabilitas yang berguna untuk menguji kesungguhan jawaban responden. Pengujian ini dilakukan dengan menggunakan aplikasi program SPSS *(Statistical Package for Social Sciences) for Windows Versi* 25.

1) Uji Validitas

Uji validitas merupakan derajat ketepatan antara data yang terjadi pada objek penelitian dengan daya yang dapat dilaporkan oleh peneliti. Dengan demikian data yang valid adalah data "yang tidak berbeda" antar data yang dilaporkan oleh peneliti dengan data yang sesungguhnya terjadi pada objek penelitian [7].

Pengujian validitas data dalam penelitian ini dilakukan secara statistik yaitu menghitung korelasi antara masing-masing pernyataan dengan skor menggunakan metode product *moment pearson correlation.* Data dinyatakan valid jika nilai r hitung yang menggunakan nilai dari Corrected Item Total Correlation > dari r tabel pada signifikansi 0,05 (5%) [2].

2) Uji Reliabilitas

Reliabilitas adalah istilah yang dipakai untuk menunjukkan sejauh mana hasil pengukuran relatif konsisten apabila alat ukur digunakan berulang kali. Untuk menguji reliabilitas kuisioner digunakan Croanbach Alpa, reliabilitas suatu instrumen memiliki tingkat realiabilitas yang tinggi apabila nilai koefesien Croanbach Alpa yang diperoleh > 0,60 [4].

### 2.4 Uji Asumsi Klasik

1) Uji Normalitas

Tujuan dilakukannya uji normalitas adalah untuk mengetahui apakah model regresi, variabel terikat dan variable bebas keduanya mempunyai distribusi normal atau tidak [2]. Data yang berdistribusi normal dalam suatu model regresi dapat dilihat pada grafik normal P-P plot, dimana bila titik-titik yang menyebar disekitar garis diagonal serta penyebarannya mengikuti arah garis diagonal, maka data tersebut dapat dikatakan berdistribusi normal.

2) Uji Multikolinieritas

Uji multikolinieritas bertujuan untuk menguji apakah model regresi ditemukan adanya korelasi antar variabel bebas (*independen*). Deteksi ada atau tidaknya multikolinieritas di dalam model regresi dapat dilihat dari besaran VIF (*Variance Inflation Factor*) dan tolerance. Regresi bebas dari multikolinieritas jika besar nilai VIF < 10 dan nilai tolerance > 0,10 [4].

3) Uji Heteroskedastisitas

Uji Heteroskedastisitas bertujuan untuk menguji apakah dalam model regresi terjadi ketidaksamaan varians dari residual satu pengamatan ke pengamatan lain [2].
Dasar pengambilan keputusan adalah sebagi berikut. Jika ada data yang membentuk pola tertentu, seperti titik-titik yang membentuk pola tertentu dan teratur (bergelombang, melebar kemudian meyempit) dan Jika tidak ada pola yang jelas serta titik-titik menyebar diatas dan dibawah angka 0 pada sumbu Y, maka tidak terjadi heterokedastisitas.

4) Uji Autukorelasi

Uji autokorelasi bertujuan untuk menguji apakah dalam model regresi linear ada korelasi antara kesalahan penggganggu pada periode t dengan kesalahan pengganggu pada periode t-1(sebelumnya) [2]. Jika terjadi korelasi, maka dinamakan ada problem autokerelasi. Adapun cara yang dapat digunakan untuk mendeteksi ada atau tidaknya autokerelasi maka digunakan uji Durbin- Watson (DW test).

### 2.5 Regresi Berganda

Evaluasi regresi pada dasarnya adalah studi mengenai ketergantunagn variabel dependen (terikat) dengan satu atau lebih variabel independen dengan tujuan untuk mengestimasi dan mempresiksi rata-rata populasi atau nilai-nilai variabel dependen berdasarkan nilai variabel independen yang diketahui [5]. Adapun rumus dari regresi berganda yang digunakan dalam penelitian ini adalah:

$$Y = \beta_0 + \beta_1 X_1 + \beta_2 X_2 + \beta_3 X_3 + \beta_4 X_4 + \beta_5 X_5 + X_6 X_6 + \beta_7 X_7 + e$$

### 2.6 Uji Hipotesis

1. Uji F

Uji F digunakan untuk mengetahui hubungan antara variabel *independen* dan variabel *dependen*, apakah variabel isi, keakuratan, bentuk, kemudahan pengunaan, ketepatan waktu, keamanan dan kecepatan berpengaruh terhadap kepuasan pengguna aplikasi Bintang *Cash &* [2].

2. Uji T

Uji T pada penelitian ini digunakan untuk mengetahui apakah variabel isi, keakuratan, bentuk, kemudahan pengunaan, ketepatan waktu, keamanan dan kecepatan berpengaruh terhadap kepuasan pengguna aplikasi Bintang *Cash & Credit* [2].

### 3. HASIL DAN PEMBAHASAN

Hasil yang akan diukur dalam penelitian ini adalah tingkat kepuasan pengguna yaitu pengguna aplikasi Bintang *Cash & Credit* dengan menggunakan metode *End User Computing Satisfaction* (EUCS) yang terdiri dari isi, keakuratan, bentuk, kemudahaan penggunaan, ketepatan waktu, keamanan dan kecepatan.

### 3.1 Analisis Kuantitatif

Skor aktual adalah jawaban seluruh responden atas kuesioner yang telah diisi. Skor ideal adalah skor atau bobot tertinggi atau semua responden diasumsikan memilih jawaban dengan skor tertinggi.

$$\text{Skor Total} = \frac{\text{Skor Aktual}}{\text{Skor Ideal}} \text{X } 100\%$$

$$\text{Skor Total} = \frac{8230}{12825} \text{X } 100\% = 64\%$$

Pengukuran setiap variabel dilakukan secara terpisah untuk mengetahui skor total dari masing-masing variabel. Diketahui bahwa tingkat kepuasan pengguna terhadap aplikasi Bintang *Cash & Credit* diperoleh nilai skor sebesar 64% yang berarti tanggapan pengguna cukup baik.

### 3.2 Analisis Statistik Deskriptif

Perhitungan *Mean* dalam analisis statistik deskriptif dimaksudkan untuk menemukan nilai rata-rata dari jawaban responden pada kuesioner.

**Tabel 1. Analisis Statistik Deskriptif**

| No | Variabel | Jumlah Mean | Hasil |
|---|---|---|---|
| 1 | Isi (*Content*) | 3.67 | Setuju |
| 2 | Keakuratan (*Accuracy*) | 3.88 | Setuju |
| 3 | Bentuk (*Format*) | 3.43 | Netral |
| 4 | Kemudahan Pengunaan (*Ease of use*) | 3.93 | Setuju |
| 5 | Ketepatan Waktu (*Timeliness*) | 3.89 | Setuju |
| 6 | Keamanan (*Security*) | 3.05 | Netral |
| 7 | Kecepatan (Speed Of Response) | 4.05 | Setuju |
| 8 | Kepuasan Pengguna (*User satisfaction*) | 4.28 | Setuju |

### 3.3 Pengujian Kualitas Data

1) Uji Validitas

Seperti yang telah dijelaskan sebelumnya bahwa uji validitas digunakan untuk mengukur sah atau tidaknya suatu kuesioner. uji validitas dilakukan dengan membandingkan nilai r hitung lebih besar dari r tabel maka item tersebut dinyatakan valid. Dalam penelitian ini untuk *degree of freedom* (df) = n – 2, dalam hal ini n adalah jumlah responden yang berjumlah 95 responden, jadi df = 95-2 = 93. Dengan tingkat signifikan 0,05 maka didapat r tabel sebesar 0,202.

**Tabel 2. Hasil Uji Validitas**

| Variabel | r hitung | r tabel | Kondisi | Kesimpulan |
|---|---|---|---|---|
| **Isi (*content*)** | | | | |
| X1_1 | 0,715 | 0,202 | r hitung > r tabel | Valid |
| X1_2 | 0,734 | 0,202 | r hitung > r tabel | Valid |
| X1_3 | 0,698 | 0,202 | r hitung > r tabel | Valid |
| X1_4 | 0,613 | 0,202 | r hitung > r tabel | Valid |
| **Keakuratan (*accuracy*)** | | | | |
| X2_1 | 0,792 | 0,202 | r hitung > r tabel | Valid |
| X2_2 | 0,846 | 0,202 | r hitung > r tabel | Valid |
| X2_3 | 0,809 | 0,202 | r hitung > r tabel | Valid |
| **Bentuk (*Format*)** | | | | |
| X3_1 | 0,820 | 0,202 | r hitung > r tabel | Valid |
| X3_2 | 0,814 | 0,202 | r hitung > r tabel | Valid |
| X3_3 | 0,784 | 0,202 | r hitung > r tabel | Valid |
| X3_4 | 0,755 | 0,202 | r hitung > r tabel | Valid |
| **Kemudahan Pengguna (*Ease of use*)** | | | | |
| X4_1 | 0,813 | 0,202 | r hitung > r tabel | Valid |
| X4_2 | 0,811 | 0,202 | r hitung > r tabel | Valid |
| X4_3 | 0,665 | 0,202 | r hitung > r tabel | Valid |
| **Ketepatan Waktu (*timeliness*)** | | | | |
| X5_1 | 0,809 | 0,202 | r hitung > r tabel | Valid |
| X5_2 | 0,836 | 0,202 | r hitung > r tabel | Valid |
| X5_3 | 0,805 | 0,202 | r hitung > r tabel | Valid |
| **Keamanan (*security*)** | | | | |
| X6_1 | 0,782 | 0,202 | r hitung > r tabel | Valid |
| X6_2 | 0,777 | 0,202 | r hitung > r tabel | Valid |
| X6_3 | 0,772 | 0,202 | r hitung > r tabel | Valid |
| X6_4 | 0,691 | 0,202 | r hitung > r tabel | Valid |
| **Kecepatan (*speed of response*)** | | | | |
| X7_1 | 0,845 | 0,202 | r hitung > r tabel | Valid |
| X7_2 | 0,808 | 0,202 | r hitung > r tabel | Valid |
| X7_3 | 0,785 | 0,202 | r hitung > r tabel | Valid |
| **Kepuasan Pengguna (*user satisfaction*)** | | | | |
| Y1 | 0,761 | 0,202 | r hitung > r tabel | Valid |
| Y2 | 0,839 | 0,202 | r hitung > r tabel | Valid |
| Y3 | 0,795 | 0,202 | r hitung > r tabel | Valid |

2) Uji Reliabilitas

Reliabilitas dilakukan untuk mengetahui konsistensi alat ukur dalam mengukur gejala yang sama. Syarat untuk menyatakan jika item itu reliabel adalah dengan melihat hasil uji reliabilitas jika setiap variabel > dari 0,60 berarti variabel tersebut reliabel.

**Tabel 3. Uji Reliabilitas**

| Variabel | Cronbach's Alpha | Keterangan |
|---|---|---|
| Isi | 0,648 | Reliabel |
| Keakuratan | 0,774 | Reliabel |
| Bentuk | 0,811 | Sangat Reliabel |
| Kemudahan pengguna | 0,685 | Reliabel |
| Ketepatan waktu | 0,764 | Reliabel |
| Keamanan | 0,755 | Reliabel |
| Kecepatan | 0,755 | Reliabel |
| Kepuasan pengguna | 0,720 | Reliabel |

**3.4 Uji Asumsi Klasik**

1) Uji Normalitas

Uji normalitas adalah pengujian tentang kenormalan distribusi data. Dari grafik terlihat bahwa nilai *plot* P-P terletak disekitar garis diagonal, *plot* P-P tidak menyimpang jauh dari garis diagonal sehingga dapat diartikan bahwa distribusi data normal regresi dapat dilihat pada gambar.

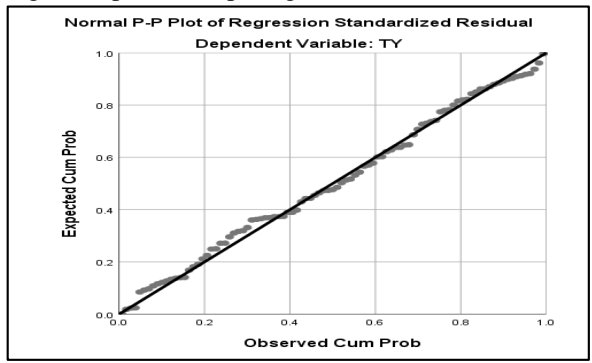

**Gambar 1. Diagram Grafik P-P Plot**

2) Uji Multikolinieritas

Uji Multikolinieritas adalah hubungan linear yang hampir sempurna atau bahkan sempurna diantara beberapa atau semua variabel *independen* dari model regresi. Bila hal ini terjadi, maka koefisien regresi berganda tidak mungkin dapat ditaksir. Untuk menguji apakah pada model regresi ditemukan adanya korelasi antar variabel bebas (*independen*). Jika nilai *tolerance value* lebih besar dari 0,10 atau *variance inflation factor* lebih besar dari 10,00 maka terjadi multikolinearitas.

**Tabel 4. Hasil Uji Multikolinieritas**

| Model | | Collinearity Statistics | |
|---|---|---|---|
| | | Tolerance | VIF |
| 1 | (constant) | | |
| | X1 | .609 | 1.642 |
| | X2 | .637 | 1.569 |
| | X3 | .580 | 1.723 |
| | X4 | .668 | 1.496 |
| | X5 | .607 | 1.649 |
| | X6 | .744 | 1.343 |
| | X7 | .695 | 1.439 |

3) Uji Heteroskedastisitas

Uji heterokedastisitas bertujuan untuk menguji apakah data model regresi terjadi ketidaksamaan *variance* dari suatu pengamatan kepenngamatan yang lain. Jika *variance* dari residual suatu pengamatan lain tetap, maka disebut heterokedastis. Cara untuk mendeteksi ada atau tidaknya heterokedastis adalah melihat grafik plot antara linai predeksi variable dependen *zpred* dengan residualnya *sresid*, dasar pengambilan keputusannya adalah sebagai berikut. Jika ada pola tertentu, seperti titik-titik yang ada membentuk pola tertentu yang teratur (bergelombang, melebar kemudian menyempit), maka mengindikasikan telah terjadi heteroskedastisitas. Jika tidak ada pola yang jelas, serta titik-titik menyebar diatas dan dibawah angka 0 pada sumbu Y, maka tidak terjadi heteroskedatisitas.

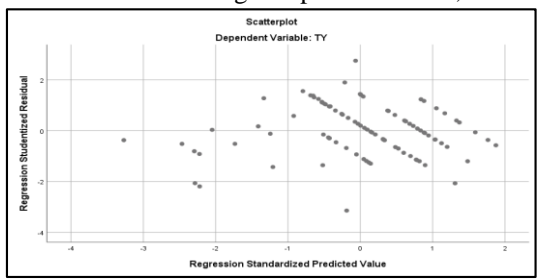

**Gambar 5. Uji Heteroskedastisitas**

4) Uji autokorelasi

Uji autokorelasi bertujuan untuk mengetahui apakah ada korelasi antara kesalahan pengganggu pada periode t dengan kesalahan pada periode t-1(sebelumnya). Dimana pengujian autokorelasi dapat dideteksi dari besarnya nilai Durbin Watson. Berikut ini merupakan petunjuk dasar pengambilan keputusan ada tidaknya autokorelasi :

**Tabel 5. Hasil uji Autokolerasi**

Model Summary[b]

| Model | R | R Square | Adjusted R Square | Std. Error of the Estimate | Durbin-Watson |
|---|---|---|---|---|---|
| 1 | .867[a] | .752 | .732 | .833 | 1.919 |

Nilai DW sebesar 1.919, nilai ini akan dibandingkan dengan nilai tabel dengan menggunakan nilai signifikansi 5%, umlah sampel 95 (n) dan jumlah variabel independen 7 (k=7), maka di tabel Durbin Watson akan didaptkan nilai Dl=1.51 dan du=1.83.

Oleh karena nilai DW 1.919 lebih besar dari batas atas (du) 1.83 dan kurang dari 4 – 1.83 (4 – du), maka dapat disimpulkan bahwa tidak terdapat autokorelasi postif atau negative.

### 3.5 Hasil Regresi Berganda

Berdasarkan perhitungan regresi antara variabel *eucs* yaitu *content, acuraccy, format, ease of use, timeliness, security, speed of use*, dengan menggunakan program SPSS 25, diperoleh hasil sebagai berikut.

**Tabel 6. Hasil Persamaan Berganda**

Coefficients[a]

| Model | | Unstandardized Coefficients | | Standardized Coefficients | T | Sig. |
|---|---|---|---|---|---|---|
| | | B | Std. Error | Beta | | |
| 1 | (Constant) | .579 | .787 | | .736 | .583 |
| | Content | .172 | .042 | .280 | 4.091 | .000 |
| | Accuracy | .109 | .051 | .143 | 2.137 | .035 |
| | Format | .064 | .043 | .104 | 1.480 | .143 |
| | ease of use | .178 | .063 | .186 | 2.845 | .006 |
| | Timeliness | .157 | .058 | .186 | 2.707 | .008 |
| | Security | .064 | .044 | .090 | 1.462 | .147 |
| | speed of response | .237 | .055 | .274 | 4.282 | .000 |

a. Dependent Variable: user satisfaction

Dari hasil persamaan regresi diatas, menunjukan bahwa nilai konstanta sebesar 0,579 artinya tanpa adanya aplikasi Bintang *Cash & Credit,* maka kepuasan pengguna hanya dinilai sebesar 0,579. Berikut uraian hasil persamaan regresi.

1) Koefisien regresi (*Content*) = 0,172 artinya apabila aplikasi Bintang *Cash & Credit* ditingkatan sebesar 0,579 maka kepuasan pengguna meningkat sebesar 0,172 dengan asumsi variabel lain dianggap konstan.
2) Koefisien regresi (*Accuracy*) = 0,109 artinya apabila apabila aplikasi Bintang *Cash & Credit* ditingkatkan sebesar 0,579 maka kepuasan pengguna meningkat sebesar 0,109 dengan asumsi variabel lain dianggap kostan.
3) Koefisien regresi (*Format*) = 0,064 artinya apabila apabila aplikasi Bintang *Cash & Credit* ditingkatkan sebesar 0,579 maka kepuasan pengguna meningkat sebesar 0,064 dengan asumsi variabel lain dianggap kostan.
4) Koefisien regresi (*Ease of use*) = 0,178 artinya apabila apabila aplikasi Bintang *Cash & Credit* ditingkatkan sebesar 0,579 maka kepuasan pengguna meningkat sebesar 0,178 dengan asumsi variabel lain dianggap kostan.
5) Koefisien regresi (*Timeliness*) = 0,157 artinya apabila apabila aplikasi Bintang *Cash & Credit* ditingkatkan sebesar 0,579 maka kepuasan pengguna meningkat sebesar 0,157 dengan asumsi variabel lain dianggap kostan.

6) Koefisien regresi (*Security*) = 0,064 artinya apabila apabila aplikasi Bintang *Cash & Credit* ditingkatkan sebesar 0,579 maka kepuasan pengguna meningkat sebesar 0,064 dengan asumsi variabel lain dianggap kostan.
7) Koefisien regresi (*Speed of response*) = 0,237 artinya apabila apabila aplikasi Bintang *Cash & Credit* ditingkatkan sebesar 0,579 maka kepuasan pengguna meningkat sebesar 0,237 dengan asumsi variabel lain dianggap kostan.

### 3.6 Uji Hipotesis

1) Uji F

Untuk mengetahui tingkat signifikan pengaruh variabel-variabel independent secara bersama-sama simultan terhadap variabel dependen dilakukan dengan menggunakan uji F yaitu dengan cara membandingkan antara F hitung dengan F tabel.

**Tabel 7. Hasil Uji F**

ANOVA[a]

| Model | | Sum of Squares | Df | Mean Square | F | Sig. |
|---|---|---|---|---|---|---|
| 1 | Regression | 182.971 | 7 | 26.139 | 37.692 | .000[b] |
| | Residual | 60.334 | 87 | .693 | | |
| | Total | 243.305 | 94 | | | |

a. Dependent Variable: user satisfaction

Berdasarkan tabel 4.19 Output Regression ANOVA diketahui nilai F hitung sebesar 37,692 dengan nilai signifikan 0,000. Untuk F tabel dapat dicari dengan melihat pada tabel f dengan signifikan 0,05 dan menentukan df1 = k-1 atau 7-1 = 6, dan df2 = n-k atau 95-7 = 88 (n = jumlah data; k=jumlah variabel independen). Di dapat F tabel adalah sebesar 2,203. Jika apabila F hitung < F Tabel maka Ho diterima dan apabila F hitung >F Tabel maka Ho ditolak.

Dapat diketahui bahwa F hitung (37,692) > F tabel (2,203) maka Ho ditolak. Jadi kesimpulannya yaitu isi (*Content*), keakuratan (*Accuracy*), bentuk (*Format*), kemudahaan pengguna (*Ease of Use*), ketepatan waktu (*Timeliness*), kemanan (*Security*) dan kecepatan (*speed of response*) secara bersama-sama berpengaruh terhadap kepuasan pengguna (*User Satisfaction*).

2) Uji T

Uji T digunakan untuk menguji pengaruh variabel independen secara persial terhadap variabel dependen. Taraf signifikan yang ditentukan adalah menggunakan nilai 0,05. Berikut adalah perhitungan uji t dari tiap variabel independen.

**Tabel 8. Hasil Uji T**

Coefficients[a]

| Model | | Unstandardized Coefficients | | Standardized Coefficients | t | Sig. |
|---|---|---|---|---|---|---|
| | | B | Std. Error | Beta | | |
| 1 | (Constant) | .579 | .787 | | .736 | .583 |
| | Content | .172 | .042 | .280 | 4.091 | .000 |
| | Accuracy | .109 | .051 | .143 | 2.137 | .035 |
| | Format | .064 | .043 | .104 | 1.480 | .143 |
| | ease of use | .178 | .063 | .186 | 2.845 | .006 |
| | timeliness | .157 | .058 | .186 | 2.707 | .008 |
| | Security | .064 | .044 | .090 | 1.462 | .147 |
| | speed of response | .237 | .055 | .274 | 4.282 | .000 |

a. Dependent Variable: user satisfaction

Dari Hasil Uji T menunjukan seberapa besar jauh pengaruh satu variabel independen secara persial dalam menerapkan variasi variabel dependen. Berikut uraian hasil uji T.

Perumusan hipotesis untuk pengambilan keputusan:
    Ho  : Tidak ada pengaruh X terhadap Y
    Ha  : Ada pengaruh X terhadap Y
Kriteria pengambilan keputusan:
    Ho  : diterima jika $t_{hitung} < t_{tabel}$
    Ha  : diterima jika $t_{hitung} > t_{tabel}$

1) *Content* (X1)

Diketahui pada tabel diatas diperoleh $t_{hitung}$ untuk variabel *content* sebesar 4,091. Sedangkan $t_{tabel}$ dapat dicari pada tabel statistik pada signifikan 0,05/2=0,025(uji 2 sisi) dengan df = n-k-1 atau 95-7-1=87 dan di dapat $t_{tabel}$ sebesar 1,988. Dengan demikian maka dapat diketahui bahwa $t_{hitung}$ (4,091) > $t_{tabel}$ (1,988) dengan nilai signifikan (*P Value*) = 0,000 < 0,05 sehingga Ho ditolak dan $Ha_1$ diterima yang berarti secara parsial (sendiri-sendiri) *content* berpengaruh terhadap kepuasan pengguna.

2) *Accuracy* (X2)

Diketahui pada tabel diatas diperoleh $t_{hitung}$ untuk variabel *accuracy* sebesar 2,137. Sedangkan $t_{tabel}$ dapat dicari pada tabel statistik pada signifikan 0,05/2=0,025(uji 2 sisi) dengan df = n-k-1 atau 95-7-1=87 dan di dapat $t_{tabel}$ sebesar 1,988. Dengan demikian maka dapat diketahui bahwa $t_{hitung}$ (2,137) > $t_{tabel}$ (1,988) dengan nilai signifikan (*P Value*) = 0,035 < 0,05 sehingga Ho ditolak dan $Ha_2$ diterima yang berarti secara parsial (sendiri-sendiri) *accuracy* berpengaruh terhadap kepuasan pengguna.

3) *Format* (X3)

Diketahui pada tabel diatas diperoleh $t_{hitung}$ untuk variabel *format* sebesar 1,480. Sedangkan $t_{tabel}$ dapat dicari pada tabel statistik pada signifikan 0,05/2=0,025(uji 2 sisi) dengan df = n-k-1 atau 95-7-1=87 dan di dapat $t_{tabel}$ sebesar 1,988. Dengan demikian maka dapat diketahui bahwa $t_{hitung}$ (1,480) > $t_{tabel}$ (1,988) dengan nilai signifikan (*P Value*) = 0,143 < 0,05 sehingga Ho diterima dan $Ha_3$ ditolak yang berarti secara parsial (sendiri-sendiri) *format* tidak berpengaruh terhadap kepuasan pengguna.

4) *Ease of use* (X4)

Diketahui pada tabel diatas diperoleh $t_{hitung}$ untuk variabel *ease of use* sebesar 2,845. Sedangkan $t_{tabel}$ dapat dicari pada tabel statistik pada signifikan 0,05/2=0,025(uji 2 sisi) dengan df = n-k-1 atau 95-7-1=87 dan di dapat $t_{tabel}$ sebesar 1,988. Dengan demikian maka dapat diketahui bahwa $t_{hitung}$ (2,845) > $t_{tabel}$ (1,988) dengan nilai signifikan (*P Value*) = 0,006 < 0,05 sehingga Ho ditolak dan $Ha_4$ diterima yang berarti secara parsial (sendiri-sendiri) *ease of use* berpengaruh terhadap kepuasan pengguna.

5) *Timeliness* (X5)

Diketahui pada tabel diatas diperoleh $t_{hitung}$ untuk variabel *timeliness* sebesar 2,707. Sedangkan $t_{tabel}$ dapat dicari pada tabel statistik pada signifikan 0,05/2=0,025(uji 2 sisi) dengan df = n-k-1 atau 95-7-1=87 dan di dapat $t_{tabel}$ sebesar 1,988. Dengan demikian maka dapat diketahui bahwa $t_{hitung}$ (2,707) > $t_{tabel}$ (1,988) dengan nilai signifikan (*P Value*) = 0,008 < 0,05 sehingga Ho ditolak dan $Ha_5$ diterima yang berarti secara parsial (sendiri-sendiri) *timeliness* berpengaruh terhadap kepuasan pengguna.

6) *Security* (X6)

Diketahui pada tabel diatas diperoleh $t_{hitung}$ untuk variabel *security* sebesar 1,463. Sedangkan $t_{tabel}$ dapat dicari pada tabel statistik pada signifikan 0,05/2=0,025(uji 2 sisi) dengan df = n-k-1 atau 95-7-1=87 dan di dapat $t_{tabel}$ sebesar 1,988. Dengan demikian maka dapat diketahui bahwa $t_{hitung}$ (1,463) > $t_{tabel}$ (1,988) dengan nilai signifikan (*P Value*) = 0,147 < 0,05 sehingga Ho ditolak dan $Ha_6$ diterima yang berarti secara parsial (sendiri-sendiri) *security* berpengaruh terhadap kepuasan pengguna.

7) *Speed of response* (X7)

Diketahui pada tabel diatas diperoleh $t_{hitung}$ untuk variabel *speed of response* sebesar 4,282. Sedangkan $t_{tabel}$ dapat dicari pada tabel statistik pada signifikan 0,05/2=0,025(uji 2 sisi) dengan df = n-k-1 atau 95-7-1=87 dan di dapat $t_{tabel}$ sebesar 1,988. Dengan demikian maka dapat diketahui bahwa $t_{hitung}$ (4,282) > $t_{tabel}$ (1,988) dengan nilai signifikan (*P Value*) = 0,000 < 0,05 sehingga Ho ditolak dan $Ha_7$ diterima yang berarti secara parsial (sendiri-sendiri) *accuracy* berpengaruh terhadap kepuasan pengguna.

## 4. KESIMPULAN

Berdasarkan hasil dan pembahasan yang telah diuraikan pada bab sebelumnya, maka dapat dibuat beberapa kesimpulan bahwa penelitian ini bertujuan untuk mengetahui pengaruh instrument EUCS (isi, keakuratan,bentuk, kemudahan pengguna, ketepatan waktu, keamanan dan kecepatan) terhadap kepuasan pengguna aplikasi Bintang *Cash & Credit* penelitian ini menggunakan data primer yang diperoleh dari kuesioner yang menggunakan pengukuran dengan skala likert. Kuesioner dibagikan kepada responden yang menggunakan aplikasi Bintang *Cash & Credit* dan kuesioner yang dibagikan adalah sebanyak 95 kuesioner. Data yang ada pada kuesioner diolah menggunakan bantuan *software* statistik SPSS 25 *for windows*.

Pertanyaan-pertanyaan yang terdapat dalam kuesioner dilakukan Pengukuran setiap variabel yang dilakukan secara terpisah untuk mengetahui skor total dari masing-masing variabel. Dapat diketahui bahwa tingkat kepuasan pengguna terhadap aplikasi Bintang *Cash & Credit* diperoleh nilai skor sebesar 64% yang berarti tanggapan pengguna cukup baik.

Penelitian ini digambarkan dalam model regresi berganda yaitu menganalisis pengaruh instrument EUCS (isi, keakuratan,bentuk, kemudahan pengguna, ketepatan waktu, keamanan dan kecepatan) terhadap kepuasan pengguna aplikasi Bintang *Cash & Credit*. Hasil penelitian menunjukkan bahwa tidak semua faktor yang bergabung dalam instrument EUCS berpengaruh terhadap kepuasan pengguna aplikasi tersebut. Dari tujuh faktor yang bergabung dalam instrumrn EUCS pada aplikasi Bintang *Cash & Credit* hanya *content* (X1), *accuracy* (X2), *ease of use* (X4), *timeliness* (X5) dan *speed of response* (X5) yang menunjukkan pengaruh signifikan terhadap kepuasan pengguna aplikasi (Y). faktor yang terdapat dalam instrumen EUCS seperti *format* (X3) dan *security* (X6) tidak menunjukkan adanya pengaruh signifikan terhadap kepuasan pengguna aplikasi Bintang *Cash & Credit*.

**DAFTAR PUSTAKA**